\documentclass[prapplied, amsmath,amssymbn,twocolumn,superscriptaddress]{revtex4-1}
\usepackage[T1]{fontenc}
\usepackage[utf8]{inputenc}

\usepackage{longtable}

\usepackage{graphicx}
\usepackage{color}
\usepackage{chemformula} 
\usepackage[T1]{fontenc} 

\usepackage[version=3]{mhchem} 
\usepackage{mathtools}
\usepackage{amssymb}

\usepackage{amsmath}

\usepackage{algorithm}
\usepackage[noend]{algpseudocode}
\newcommand{\myalgorithm}{
\begin{algorithm}[H]
\small
\begin{algorithmic}[1]
\State Set the High level method and the software to be used (here B2PLYP/def2tzvpp using Gaussian)
\State Optimize geometry using the high level method
\State Compute harmonic modes using the high level method
\State Generate  $3N\times8$ geometries for ANN training potential
\For{each geometry} \Comment{This loop can be run in parallel}
\State \hspace{\algorithmicindent} Compute B2PLYP/def2tzvpp energies and forces analytically.
\EndFor
\State Build the ANN potential by training on the energies and forces of geometries generated at previous step. Note that ANN forces are computed analytically.
\State Generate all geometries needed to calculate the cubic and quartic derivatives using the ANN potential
\For{each geometry} \Comment{This loop can be run in parallel}
\State \hspace{\algorithmicindent} Compute energy
\EndFor
\State Using the calculated energies, compute cubic and quartic energy derivatives
\State Using these derivatives, compute anharmonic frequencies by GVPT2 method
\end{algorithmic}
\end{algorithm}
}

\begin{document}

\title{Combining quantum mechanics and machine-learning calculations for 
anharmonic corrections to vibrational frequencies} 

\author{Julien Lam}
\affiliation{Center for Nonlinear Phenomena and Complex Systems, Code Postal 231, Universit\'e Libre de Bruxelles, Boulevard du Triomphe, 1050 Brussels, Belgium}
\email{julilam@ulb.ac.be}

\author{Saleh Abdul-Al}
\affiliation{Lebanese International University, Bekaa, Lebanon and International University of Beirut, Beirut, Lebanon}
\author{Abdul-Rahman Allouche}
\affiliation{ Institut Lumi\`ere Mati\`ere, UMR5306 Universit\'e Lyon 1-CNRS, Universit\'e de Lyon, 69622 Villeurbanne Cedex, France }
\email{abdul-rahman.allouche@univ-lyon1.fr}


\begin{abstract}
Several methods are available to compute the anharmonicity in semi-rigid molecules. However, such methods are not routinely employed yet because of their large computational cost, especially for large molecules. The potential energy surface is required and generally approximated by a quartic force field potential based on \textit{ab initio} calculation, thus limiting this approach to medium-sized molecules. We developed a new, fast and accurate hybrid Quantum Mechanic/Machine learning (QM//ML) approach to reduce the computational time for large systems.  With this novel approach, we evaluated anharmonic frequencies  of 37 molecules thus covering a broad range of vibrational modes and chemical environments. The obtained fundamental frequencies reproduce results obtained using B2PLYP/def2tzvpp with a  root-mean-square deviation (RMSD) of 21 cm$^{-1}$ and experimental results  with a RMSD of 23 cm$^{-1}$. Along with this very good accuracy, the computational time with our hybrid QM//ML approach scales linearly with N while the traditional full ab initio method scales as N$^2$, where N is the number of atoms.
\end{abstract}

\maketitle

\section{Introduction}

Recent advances in vibrational spectroscopy originate from progress at both experimental and theoretical levels. On the one hand, the development of ultrafast lasers has allowed for probing molecular dynamics on extremely short time scales\cite{Mayda2019Apr,Kondo2006Oct}. On the other hand, computational spectroscopy has benefited from novel quantum mechanics calculations whose accuracy is comparable to experimental measurements\cite{Christiansen2012Apr,Gruenbaum2013Jul,Joutsuka2016Oct,Bistafa2019Aug,Yang2017Nov,Medders2015Mar}. In this research field, the combining role of experiments and simulations is two-fold: Experiments help to assess the accuracy of the numerical calculations, which in turn, enable for disentangling the complexity of the experimental measurements\cite{Schindler2017Oct,Schindler2018Oct,Gaynor2018Nov,Kozubal2019Jun,Brunken2019Sep}. To obtain additional progress in computational spectroscopy, the current challenge is to carry out calculations for large molecules while maintaining the accuracy obtained for the smaller ones\cite{Hirshberg2017Mar,Ivani2010Jul,Barnes2016,Roy16,Yagi19,keceli11}. Such glass ceiling proves to be even more difficult to break when it comes to computing the anharmonic corrections to vibrational frequencies\cite{Hirshberg2017Mar,Ivani2010Jul,Barnes2016,Roy16,Yagi19,keceli11}. 

Indeed, anharmonicity of the electronic potential is often neglected and taken into account a posteriori using corrective factors that are empirically determined to match experimental measurements. More rigorously, explicit modeling of anharmonicity is also carried out using several methods including VSCF\cite{Bowman78}, VSCF-PT2\cite{Yagi00}, cc-VSCF\cite{Jung96}, VCI-P\cite{Carbonniere10,Erba19-1,Erba19-2}, VT2\cite{Nielsen51}, GVPT2\cite{Martin95,Barone2005Jan,Barone2014Jan,Cornaton16,Charmet18}.
Because of the required high-level of ab-initio calculations, the current calculations are computationally demanding for large molecules, in particular with double-hybrid functionals which consist on mixing hybrid functionals with an additional second-order perturbation theory correlation.

Alternatively, instead of such a full ab-initio calculation, several studies have been suggested. For solvated molecules, the computing time can be reduced using a hybrid QM//MM potential\cite{Yagi19}. Other methods remove many derivatives terms, using only the 2-mode coupling quartic force constants\cite{keceli11}. But, the most popular approach is a hybrid one where the geometric structure optimization and harmonic calculations are computed at a high level of theory as the DFT theory and anharmonic corrections are obtained with semi-empirical methods.\cite{Roy16}. More recently, we proposed to use molecular mechanic (MM) potential and in particular, the MMFF94\cite{Halgren96} method which represents an even lower level of theory\cite{Barnes2016}. The success of such a hybrid approach is two-fold. On the one hand, the accuracy is comparable to full \textit{ab initio} calculations, and on the other hand, it allows for a considerable reduction of computational costs. However, MMFF94 method can only be employed for organic molecules and an alternative approach must be derived.

Machine-learning methods have been recently employed to bridge the gap between molecular mechanic potentials and \textit{ab initio} calculations by providing a framework to reach the accuracy of the latter while maintaining a low computational cost\cite{Behler2016Nov,Schmidt2019Aug,Hill2016May}. Four classes of methods are usually considered: (1) Gaussian approximation potentials\cite{Bartok2010Apr,Bartok2015Aug}, (2) Kernel Ridge Regression\cite{Hu2018May,Vu2015Aug,Rupp2012Jan}, (3) Sparse Linear Regression\cite{Seko2015Aug,Seko2014Jul,Takahashi2018Jun,Takahashi2017Nov} and (4) Artificial Neural Network (ANN)\cite{Behler2007Apr,Behler2011Oct,Artrith2016Mar}. Recent comparisons between these different methods have also been carried out\cite{Li2018Dec,Yunxing2019}. 

In this context, we propose the use of  machine-learning techniques and in particular ANN potential. This way, our novel approach overcomes the limitations of the previous one and is now versatile enough to compute anharmonic corrections to vibrational frequencies in organic, as well as inorganic molecules. Such hybrid approach has already been investigated in the literature and tested on few molecules but required the use of molecular dynamics simulations\cite{Gastegger2017Sep}. In this paper, we wish to extend the methodology by using only static calculations. We will first describe the methodology with particular details on the choices that were made for the ANN parameters and the benchmarking molecules. Then, we will compare the obtained results with experiments and also more advanced \textit{ab initio} calculations.

\section{Methods}

\subsection{The hybrid Quantum mechanics/Machine learning approach}

Anharmonic corrections to vibrational frequencies are computed within the explicit framework of the generalized second-order vibrational perturbation theory (GVPT2)\cite{Barone2005Jan,Barone2014Jan}. We will only briefly outline key features of the approach before describing more thoroughly the choices that were made for our own implementation. For a given molecule of $N$ atoms, the potential energy surface is a function of the normal coordinates denoted $\bold{Q}$. There are $f$ normal coordinates with $f=3N-6$ for a non-linear molecule and $f=3N-5$ otherwise. Within GVPT2, the potential energy surface (PES) is approximated by a quartic force field using:

\begingroup
\scriptsize

\[
V(\bold{Q}) = V_0 + V_1(\bold{Q}) + V_2(\bold{Q}) + V_3(\bold{Q}) \\
\]
\[
V_1(\bold{Q}) = \sum \limits_{i=1}^{f} { \frac{1}{2} h_i Q_i^2 + \frac{1}{6} t_{iii} Q_i^3 + \frac{1}{24} u_{iiii} Q_i^4}\\
\]
\[
V_2(\bold{Q}) = \sum \limits_{ij, i \neq j}^{f} {\frac{1}{2} t_{ijj} Q_i Q_j^2 + \frac{1}{6} u_{ijjj} Q_i Q_j^3} + \sum \limits_{ij,i<j}^{f} {\frac{1}{4} u_{iijj} Q_i^2 Q_j^2}\\
\]
\[
V_3(\bold{Q}) = \sum \limits_{ijk, i \neq j<k}^{f} { t_{ijk} Q_i Q_j Q_k} + \sum \limits_{ijk,i \neq j<k}^{f} {\frac{1}{2} u_{iijk} Q_i^2 Q_j Q_k}\\
\]
\endgroup

where $V_0$, $h_i$, $t_{ijk}$ and $u_{ijkl}$ denote respectively the energy and its second-, third-, and fourth-order derivatives with respect to the normal coordinates at the equilibrium geometry,  calculated through numerical differentiations of the energy (see Yagi et al.\cite{Yagi04} ). From there, computing vibrational frequencies requires calculating those derivatives. In our implementation, the second derivatives are obtained directly using harmonic frequencies from high-level DFT calculations. Then, the third and the fourth derivatives are computed using a fitted potential based on a feed-forward neural network. As such, the energy for each atom depends on symmetry functions which measure the local environment. The relationship between these symmetry functions and the atomic energy is described by a neural network which consists of ten neurons and two hidden layers. 
In particular, we used the Behler and Parrinello\cite{Behler2011Feb} approach which is based on the assumption that the total potential energy $V^{pot}$ can be written as the sum of individual atomic
contributions: $  V^{pot} = \sum_{j=1}^{N} E_j(\mathbf {G_j}) $ where N is the number of atoms. The atomic energy $E_j$ is calculated from neural networks with multiple layers of nodes.
The neurons of input layer are populated with sets of real number $\mathbf G_j$, called symmetry functions to caracterise the local environnment of atom j. The analytical experession of $\mathbf G_j$ and the parameters for the symmetry functions are given in supporting information. Then, from the vector of inputs $y_0 = \mathbf G_j$ in the input layer, the output of a node i in layer k can be computed according to  : 
$y_i^k = f_a(w^k_{0i} + \sum_{m=1}^{n_l} w^{lk}_{mi} y_m^l$) where $y^k_i$ is the value of neuro i in layer k, $w^k_{0j}$ is the associated bias, $w^{lk}_{mi}$ is the weight connecting $y^k_i$ to $y_m^l$ and $n_l$ is the number of neuron in layer $l = k-1$. $f_a$ is the acivation function, used as a hyperbolic tangent in our work.

The parameterization of the neural network components is carried out using the Kalman Filter optimization method\cite{Singraber2019Apr} on energy and forces obtained with training geometries. The n2p2 Neural Network Potential Package\cite{Singraber2019Jan} was used to build the ANN potential\cite{Singraber2019Jan, Singraber2019Apr}

We performed two types of quantum mechanics calculations:  (1) geometry optimization to obtain equilibrium positions of the atoms and the harmonic frequencies which is also necessary  with any full ab initio approach and (2) additional single-point (energy and forces) calculations around  the equilibrium positions to train the neural network potential. In particular, we take in consideration random displacements of each atom from the equilibrium position in the three Cartesian directions by $\pm0.01$, $\pm0.05$, $\pm0.2$ and $\pm0.3$\AA\,which leads to $3N\times8$ geometries per molecule. With such displacements, we make sure that atoms are sufficiently displaced to sample the anharmonicity of the PES. We used 95\% of the data sets for the training of the neural network and 5\% for the testing, and the root mean squared error is less than 0.5 meV for energy and 30 meV/\AA\, for forces. Altogether, the quantum-mechanics calculations are carried out with the B2PLYP\cite{Grimme06}/def2tzvpp\cite{Weigend05} method as implemented in the Gaussian09 package\cite{G09}. We chose to perform our calculation with the double hybrid B2PLYP method because it is the most accurate method that can still be employed to compute anharmonic frequencies with a molecule made of tens of atoms\cite{Biczysko2010Jul}. As such, we managed to compare our hybrid results with full ab initio calculations. However, in our approach, the database can very well be constructed with any other ab initio method (DFT, MP2, Coupled-Cluster, ...)  and we foresee the use of our hybrid approach with more accurate ab initio methods. Finally, the anharmonic corrections are performed using our iGVPT2\cite{iGVPT2} code, linked to the n2p2 library to compute ANN energies. See Figure \ref{figure:algo} for an algorithmic representation of the entire hybrid method. 

Altogether in our approach, only $24N$ DFT calculations of energies and forces per molecules are required  to construct the quartic force field while for a full DFT approach, it is necessary to perform $(3N-6)(3N-5)$ calculations of energies and forces, as implemented in the Crystal program\cite{Erba2019Jun,Erba2019Jun2} and $6N-11$ Hessian calculations as implemented in Gaussian\cite{Barone2005Jan}. However, the time needed to perform a Hessian calculation, even analytically, is more expensive than a single-point energy+forces calculation. If the Hessian matrix is computed numerically from forces, $2(3N-6)$ energies and forces calculations are needed for each Hessian and in overall $2(6N-11)(3N-6)$ energies and forces calculations are then required to build the quartic force field.

\begin{figure}
\footnotesize{
    \myalgorithm
}
    \caption{Hybrid B2PLYP//ANN approach-pseudocode}
    \label{figure:algo}
\end{figure}

For simplicity,  in the next sections, we refer to :
\begin{itemize}
\item Full B2PLYP as a GVPT2 calculation where all the values are calculated using the B2PLYP/def2tzvpp method via Gaussian software.
\item Hybrid B2PLYP/ANN as our hybrid method where the B2PLYP/def2tzvpp was used to compute the harmonic modes while the cubic and fourth ones are computed using the ANN potential.
\item Full ANN as calculation where all derivatives are calculated using ANN potential via a custom-made interface to n2p2 library. The fundamental frequencies are calculated using GVPT2 method.
\end{itemize}

\subsection{Reference data}

The hybrid Quantum mechanics/Machine learning approach is tested on 37 molecules:
Water (\ce{H2O}), Krypton difluoride (\ce{KrF2}), Carbonyl selenide (\ce{OCSe}), Sulfur dioxide (\ce{SO2}), selenium dioxide (\ce{SeO2}), Carbon diselenide (\ce{CSe2}), Titanium dioxide (\ce{TiO2}), Ammonia (\ce{NH3}), chlorine trifluoride (\ce{ClF3}), Formaldehyde (\ce{H2CO}), Hydrogen peroxide (\ce{H2O2}), Gallium trichloride (\ce{GaCl3}), Aluminum chloride difluoride (\ce{AlF2Cl}), aluminum dichloride fluoride (\ce{AlFCl2}), Zinc methylene (\ce{ZnCH2}), Methane (\ce{CH4}), Bromoform (\ce{CHBr3}), silicon tetrabromide (\ce{SiBr4}), Titanium tetrachloride (\ce{TiCl4}), Carbon tetrabromide (\ce{CBr4}), Formic acid (\ce{CH2O2}), Methanimine (\ce{CH2NH}), Methyl Alcohol (\ce{CH3OH}), Methane selenol (\ce{CH3SeH}), Vinyl bromide (\ce{C2H3Br}), Chlorethene (\ce{CH2CHCl}), Selenium hexafluoride (\ce{SeF6}), 1,2-Dibromoethane (\ce{C2H4Br2}), methyl germane (\ce{GeH3CH3}), Ethyl bromide (\ce{C2H5Br}), Cyclopropane (\ce{C3H6}), Propylene oxide (\ce{C3H6O}), pyruvic acid (\ce{C3H4O3}), benzene (\ce{C6H6}), Thiophene (\ce{C4H8S}), Dimethyl sulfate (\ce{C2H6O4S}) and Naphthalene (\ce{C10H8}).

All of these molecules were specially chosen because they span over a large chemical space and because  their vibrational frequencies were experimentally measured (taken from NIST Database)\cite{NIST}. Altogether, our benchmarking is made on 407 experimental fundamental frequencies available to the 37 molecules. For molecules where it is possible to compute the full B2PLYP/def2tzvpp anharmonic corrections to vibrational frequencies, we used them as a second way to assess the accuracy of our approach. For this second comparison, we worked with 371 frequencies available to the 34 molecules. Three molecules could not be investigated with the full B2PLYP/def2tzvpp calculations because of large computational cost (\ch{C10H8}) or because of high symmetry of the molecule whose anharmonic calculations can not be performed with the employed version of Gaussian (\ch{SeF6} and \ch{C6H6}).  

\section{Benchmarks}

For each molecule of our data set, after optimization of the geometry using B2PLYP/def2tzvpp method, the $3N\times8$ geometries (where $N$ is the number of atoms) are generated. Their B2PLYP/def2tzvpp energies and forces are calculated and used to build the corresponding ANN potential. Fitting results are given in Table S3 of supporting information. The RMSD for forces used in training varies from 5.0 to 23.0 meV/\AA\, with an averaged value of 14.8 meV/\AA.  That of energies varies from 0.1 to 0.6 meV with an averaged value of 0.21 meV. This excellent result proves that our symmetry functions with only 10 neurons in 2 hidden layers can reproduce, with excellent accuracy the potential energy surface around the equilibrium geometry.

To validate the performance of our hybrid B2PLYP/ANN GVPT2 approach, we carried out the calculation of the fundamental frequencies on all molecules of our data set.  To compare the accuracy of our Hybrid method to that of the standard GVPT2 approach (full B2PLYP), we calculated, when possible the fundamental frequencies of molecules using this method. Full B2PLYP is known as a very accurate method to study small- and medium-sized molecules but the computational cost rapidly becomes prohibitive for large molecules\cite{Biczysko2010Jul}. All calculated and experimental frequencies are given in Table S4 in supporting information.

\subsection{Assessment of the hybrid approach by comparison with full B2PLYP one}

\begin{figure}[!h!]
\includegraphics[width=\columnwidth]{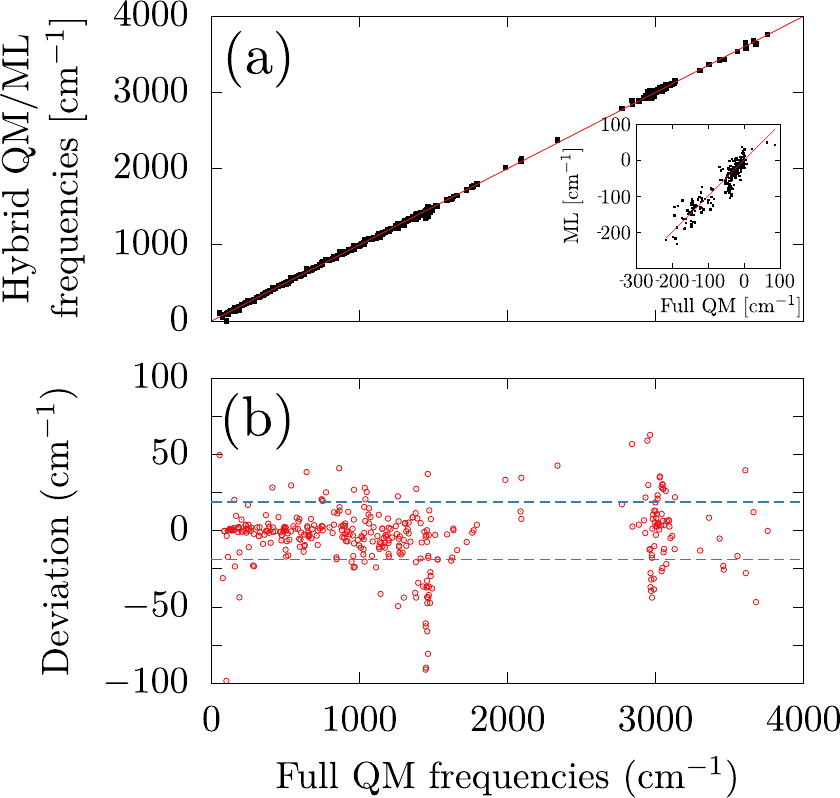}
\caption{(a) Scatter plots for the fundamental frequencies/cm$^{-1}$ and the anharmonic corrections in inset. Black points are for present data while the red line corresponds to the ideal case. (b) Deviation between the Hybrid QM/ML frequencies and the full QM frequencies. Blue dotted lines correspond to  deviations of $\pm 20$ cm$^{-1}$.}
\label{ScaterPlotHybridVsB2PLYP}
\end{figure}

Frequencies obtained from our hybrid model are compared to results obtained from the full B2PLYP approach in Figure \ref{ScaterPlotHybridVsB2PLYP}.a. The correlation coefficients R$^2$ between our hybrid ANN and full B2PLYP calculations are equal to 0.9996 for fundamental frequencies and 0.934 for anharmonic corrections. Averaging through our entire benchmark set (Table \ref{table:tableErrors}), we obtained a root mean squared deviation (RMSD) of $21$ cm$^{-1}$ for all the frequencies and it becomes $20$ cm$^{-1}$ and $24$ cm$^{-1}$ when respectively restricted to only the low and high frequencies. The unsigned absolute error (AUE) are about 13 cm$^{-1}$, 11 cm$^{-1}$ and 19 cm$^{-1}$ for all, low and high frequencies respectively. In addition, we also plotted in Figure \ref{ScaterPlotHybridVsB2PLYP}.b the corresponding deviation as a function of the full QM frequencies. Altogether, these results reflect the overall very good quality of the agreement between the hybrid and the full B2PLYP approaches. 

\begin{table}
\caption{Statistical errors in $cm^{-1}$ using all frequencies, low frequencies ($<=$2000 cm$^{-1}$) and High ones($>$2000 cm$^{-1}$). In the column denoted "Reference", the line "B2PLYP" and "Experiment" correspond to cases where statistical errors are computed using respectively B2PLYP and Experimental frequencies\cite{NIST} as reference.}
\label{table:tableErrors}
\scriptsize{
\setlength{\tabcolsep}{2pt}
\begin{tabular}{ |lllrrrr|}
\hline
                 All &          Frequencies &                      &                      &                      &                      &                      \\ 
              Method &                 Type &            Reference &                 RMSD &                  AUE &                  ASE &                 UMAX \\ 
              Hybrid &          Fundamental &               B2PLYP &                20.54 &                13.11 &                 3.22 &                96.74 \\ 
              Hybrid &          Fundamental &           Experiment &                22.93 &                16.19 &                 0.56 &               117.91 \\ 
              B2PLYP &          Fundamental &           Experiment &                19.80 &                13.07 &                 -3.82 &                98.51 \\ 
                 ANN &          Fundamental &               B2PLYP &                21.99 &                13.97 &                 2.42 &               174.10 \\ 
                 ANN &          Fundamental &           Experiment &                23.11 &                16.32 &                -1.00 &                86.93 \\ 
                 ANN &             Harmonic &               B2PLYP &                18.15 &                11.88 &                -2.47 &                98.47 \\ 
\hline
                 Low &          Frequencies &                      &                      &                      &                      &                      \\ 
              Method &                 Type &            Reference &                 RMSD &                  AUE &                  ASE &                 UMAX \\ 
              Hybrid &          Fundamental &               B2PLYP &                19.60 &                11.62 &                 5.08 &                98.51 \\ 
              Hybrid &          Fundamental &           Experiment &                20.80 &                14.26 &                 3.69 &               117.91 \\ 
              B2PLYP &          Fundamental &           Experiment &                16.18 &                10.35 &                 -1.15 &                85.86 \\ 
                 ANN &          Fundamental &               B2PLYP &                23.11 &                14.03 &                 3.15 &               174.10 \\ 
                 ANN &          Fundamental &           Experiment &                21.79 &                15.16 &                 1.27 &                85.55 \\ 
                 ANN &             Harmonic &               B2PLYP &                19.10 &                12.21 &                -3.72 &                98.47 \\ 
\hline
                High &          Frequencies &                      &                      &                      &                      &                      \\ 
              Method &                 Type &            Reference &                 RMSD &                  AUE &                  ASE &                 UMAX \\ 
              Hybrid &          Fundamental &               B2PLYP &                23.66 &                18.54 &                -3.53 &                62.88 \\ 
              Hybrid &          Fundamental &           Experiment &                29.40 &                23.23 &               -10.85 &                84.38 \\ 
              B2PLYP &          Fundamental &           Experiment &                28.60 &                21.95 &                 -12.52 &                96.74 \\ 
                 ANN &          Fundamental &               B2PLYP &                17.29 &                13.77 &                -0.27 &                47.62 \\ 
                 ANN &          Fundamental &           Experiment &                27.40 &                20.55 &                -9.25 &                86.93 \\ 
                 ANN &             Harmonic &               B2PLYP &                13.82 &                10.60 &                 2.38 &                37.14 \\ 
\hline
\end{tabular}
}
\end{table}

\begin{figure}[!h!]
\includegraphics[width=\columnwidth]{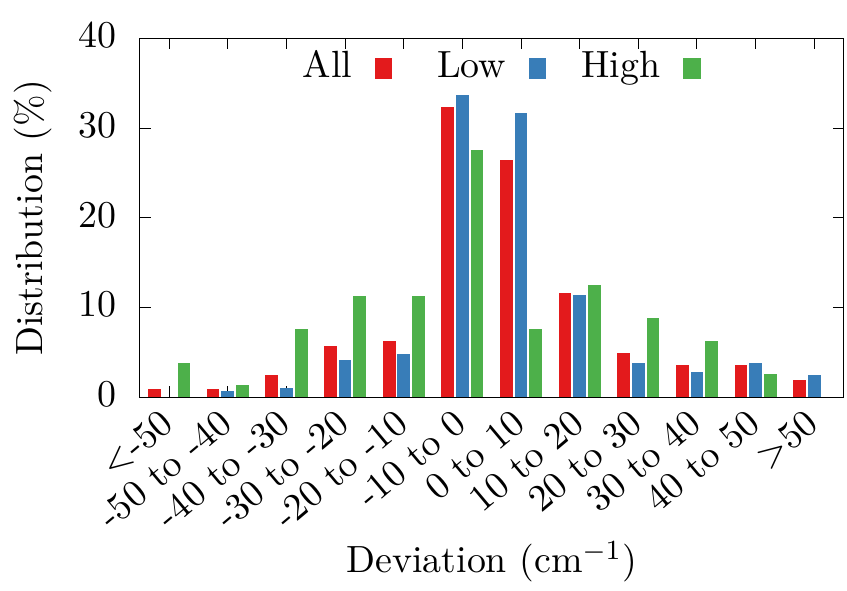}
\caption{Distribution of the vibrational frequency deviation between hybrid calculations and full \textit{ab initio} calculations.}
\label{VsSimu}
\end{figure}

More precisely, Fig.\ref{VsSimu} shows the distribution of deviation between our hybrid calculations and full B2PLYP ones. Most frequencies are obtained with a deviation smaller than $10$ cm$^{-1}$. The distribution is peaked around 0, where 59\% of frequencies are predicted with a deviation between $-10$ to $10$ cm$^{-1}$, 77\% between $-20$ and $20$ cm$^{-1}$ and 81\% between $-30$ and $30$ cm$^{-1}$.

Following Fig.\ref{VsSimuPerMol}, it appears that our results depend on the considered molecule. In particular, the largest molecules including Cyclopropane(\ce{C3H6}) and Methyl German \ce{GeH3CH3} exhibit the largest error which is due to the higher difficulty in obtaining a good description of anharmonic effect on bend scissor HCH modes. This error is not due to the number of neurons or to the Group Symmetry Function. Indeed, the forces are reproduced with a small RMS (11 meV/\AA\, for \ce{C3H6} and 14 meV/\AA\, for \ce{GeH3CH3}). Despite this specific case, the averaged RMSD on over all modes of all molecules remains very small.

Altogether, our results are consistent with previous precision obtained using MMFF94\cite{Barnes2016}. However, the advantage of this novel approach is the large versatility as virtually any molecules can be considered.

\begin{figure}[!h]
\includegraphics[width=\columnwidth]{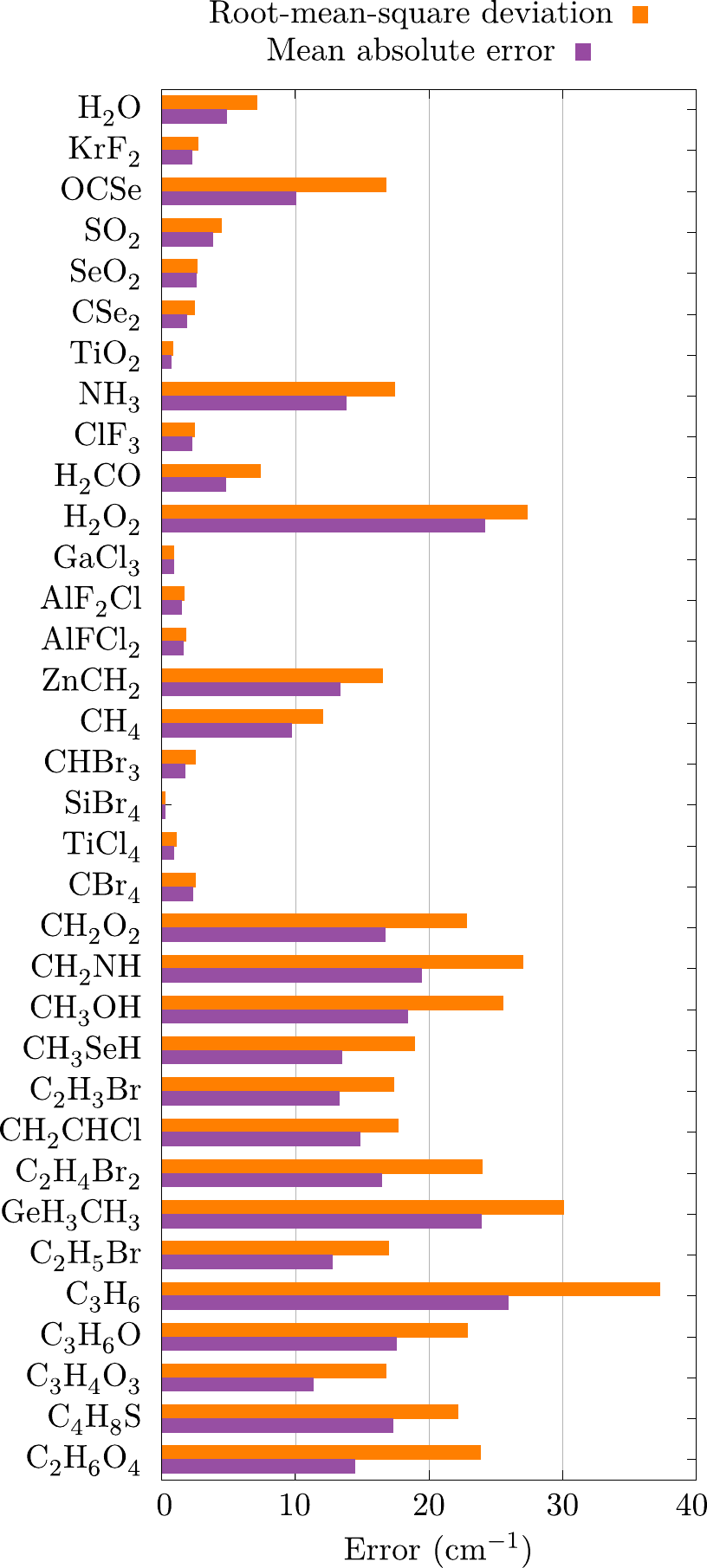}
\caption{Error in vibrational frequency obtained by hybrid calculations relative to full \textit{ab initio} calculations for each molecule.}
\label{VsSimuPerMol}
\end{figure}

\subsection{Assessment of the hybrid approach by comparison with experiment}

 \begin{figure}[h!]
\includegraphics[width=\columnwidth]{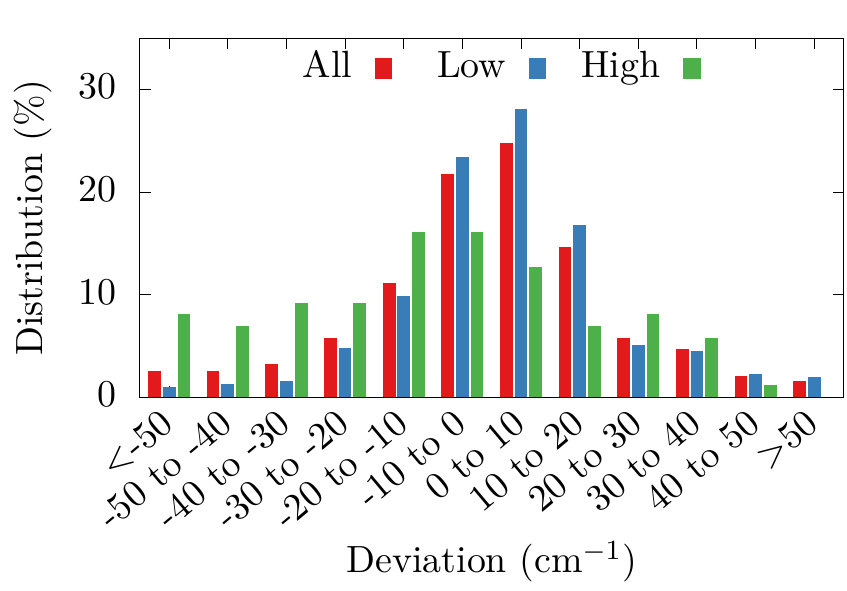}
\caption{Distribution of the vibrational frequency deviation between hybrid calculations and experimental measurements.}
\label{VsExp}
\end{figure}

Fig.\ref{VsExp} shows the deviation between our hybrid calculations and experimental measurements. 
Most frequencies are obtained with a deviation smaller than $10\,$cm$^{-1}$ with a distribution peaked around 0. Averaging through our entire benchmark set, we obtained a root mean squared deviation (RMSD) of $23\,$cm$^{-1}$ for all the frequencies(Table \ref{table:tableErrors}) and it becomes $21\,$cm$^{-1}$ and  $29\,$cm$^{-1}$ when respectively restricted to only the low and high frequencies. The mean absolute errors are $16\,$cm$^{-1}$, $14\,$cm$^{-1}$ and $23\,$cm$^{-1}$  for respectively all the frequencies, only the low and only the high frequencies. This error can originate from two sources: (1) Dataset used to fit the ANN potential to the ab initio method (B2PLYP) and (2) The method (GVPT2) used to compute the fundamental frequencies. To distinguish between these two,  the frequencies obtained with the full B2PLYP approach are compared with experimental ones which leads to a RMSD equal to  20, 16 and $29\,$ cm$^{-1}$ for respectively all, low and high frequencies. The full ab initio calculations lead to an agreement with experiments that is similar to the hybrid approach. Altogether, this demonstrates that the ANN potential does not really provide any additional error.  

\begin{figure}[h!]
\includegraphics[width=\columnwidth]{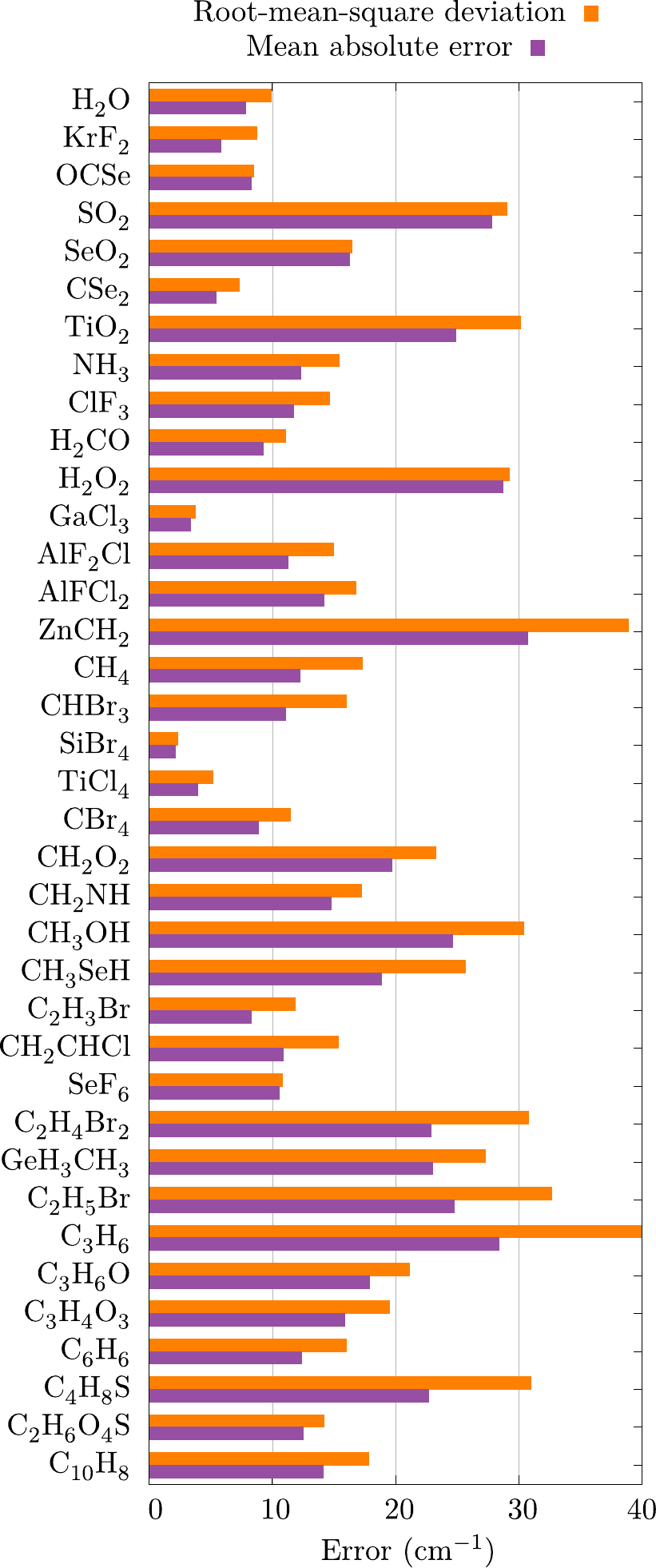}
\caption{Error in vibrational frequency obtained by hybrid calculations and experimental measurements for each molecule.}
\label{VsExpPerMol}
\end{figure}

Fig.\ref{VsExpPerMol} shows that the errors depend on the considered molecules. The large error obtained for Cyclopropane is due to the frequency mode discussed in the previous section. In addition, we observe a large error for \ce{ZnCH2}. This error is due to B2PLYP/GVPT2 approach not to ANN potential. Indeed the RMSD obtained with full B2PLYP/GVPT2 is about $40$ cm$^{-1}$. The CH2 wagging mode cannot be reproduced correctly with this approach. We note that there is a Fermi resonance between CH2 wagging and CH2 scissor modes, treated, in GVPT2 approach, variationally after removal of resonant term using Martin et al. criteria\cite{Martin95}. A VCI calculation should be more appropriate to study the anharmonic effect on this molecule. An error in the experimental value cannot be excluded either. Indeed, the $543.8$ cm$^{-1}$ of CH2 wagging mode in CH2 seems to be too small compared to the same mode in other molecules.

\subsection{Assessment of full ANN approach}
\begin{figure}[!h!]
\includegraphics[width=\columnwidth]{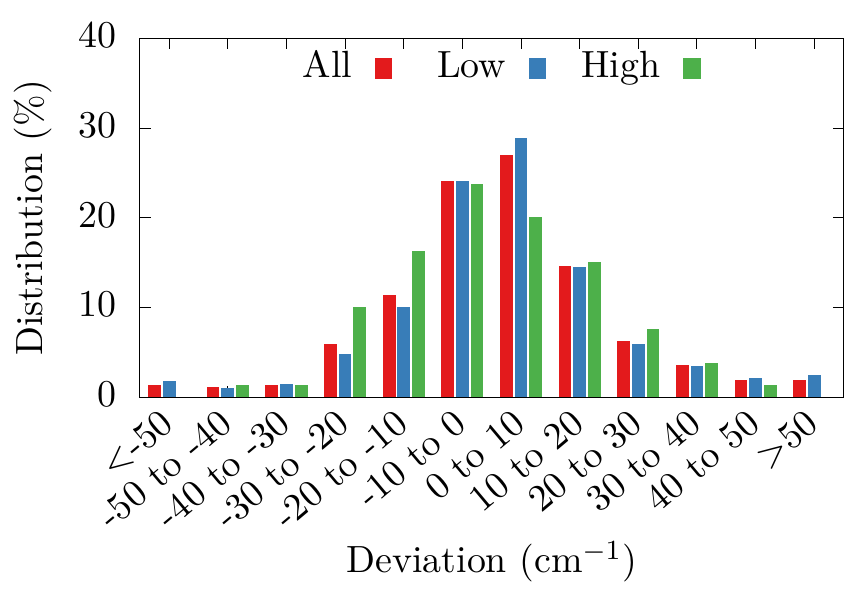}
\caption{Distribution of the vibrational frequency deviation between full ANN calculations and full \textit{ab initio} calculations.}
\label{ANNVsSimu}
\end{figure}

\begin{figure}[h!]
\includegraphics[width=\columnwidth]{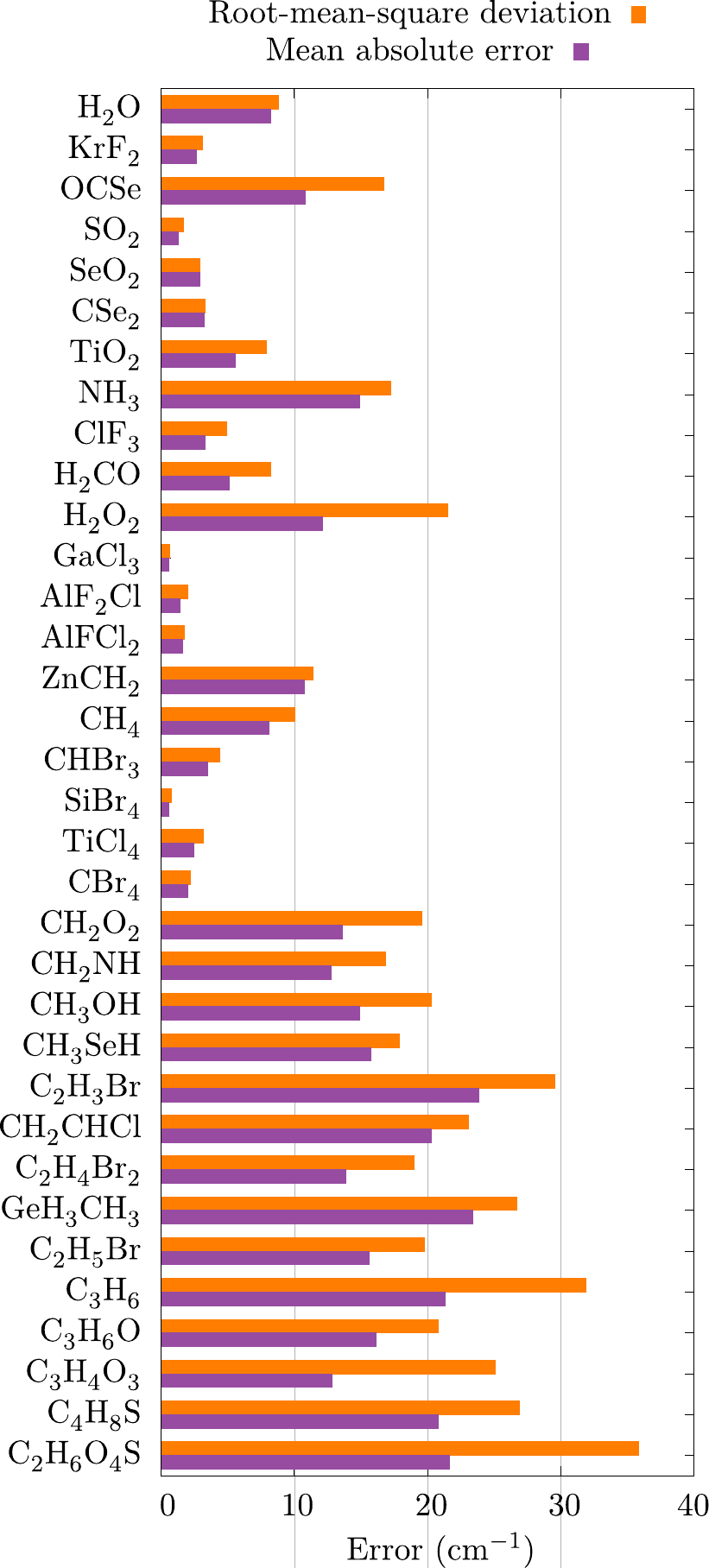}
\caption{Error in vibrational frequency obtained by full ANN calculations and full \textit{ab initio} ones.}
\label{ANNVsSimuPerMol}
\end{figure}

We have seen above that hybrid B2PLYP/ANN approach gave very good results as compared with the full B2PLYP approach. Therefore, we further considered an alternative approach, where equilibrium geometry, second, cubic and quartic derivatives are all calculated using the PES obtained with the ANN potential. When comparing with the full B2PLYP approach, 
RMSD for fundamental frequencies is equal to $22$, $23$ and $17$ cm$^{-1}$ for respectively all, low and high frequencies, and the RMSD for harmonic frequencies is equal to $18$, $19$ and $14$ cm$^{-1}$ for respectively all, low and high frequencies (Table 1). This shows that our full ANN approach also leads to very good results as compared with B2PLYP(See Figures \ref{ANNVsSimu} and \ref{ANNVsSimuPerMol} ). The comparison between Figures 7 and 2 shows very similar deviations in the vibrational frequency with slightly better results for Hybrid B2PLYP/ANN approach. Similarly, the comparison of our full ANN approach with the experimental results shows very good agreement.
While the RMSD of full ANN approach is similar to that of our hybrid B2PLYP/ANN one, this full ANN approach leads, in general, to slightly less accurate results which arises from the errors already obtained in the ANN calculation of few very-low harmonic frequencies. Such error could be coped with by increasing the number of data in the training set which naturally increases the computational cost. Altogether, the full ANN approach remains a good alternative to hybrid B2PLYP/ANN one, especially for very large molecules where the harmonic B2PLYP calculation can take very large computational time.

\subsection{Timings}
\begin{figure}[h!]
\includegraphics[width=\columnwidth]{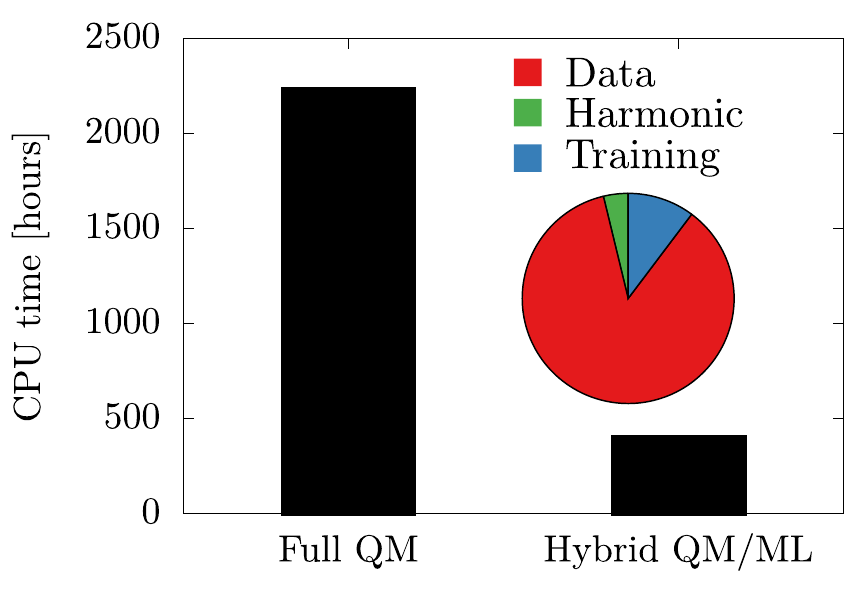}
\caption{CPU time/hours for full B2PLYP and hybrid B2PLYP/ANN calculations. Time for each step of hybrid B2PLYP/ANN ones.}
\label{TimingRealTime}
\end{figure}

Regarding the computational timing, we take \ce{C4H8S}, relatively one of the large molecules in our benchmark set, as an example. As illustrated in Figure \ref{TimingRealTime}, when using the full B2PLYP treatment, it took 2240 hours in CPU time and 72 hours in real-time on 40 cores. Calculations using the hybrid B2PLYP/ANN was considerably reduced as it took only 400 hours in CPU time and 7 hours in real-time (using 100 cores for Data and only 8 cores for the training part). The main advantage here is that for each molecule, calculations of the distorted structures were run in parallel. It is evident that the computational time should grow with the number of atoms.  In the hybrid calculation, the time grows as $3N \times 8$ while it grows as $\approx(3N)^2$ for a full \textit{ab initio} calculation.\cite{Barnes2016}. Hence, beside its very good accuracy, the hybrid calculation approaches is ten times faster than full ab initio methods. This significant reduction in the computational time enables us to carry out the anharmonic corrections to vibrational frequencies of large molecules, which was not feasible before. 

\subsection{Effect of data sets}
The main goal in this work was to construct an accurate PES through a fast method. This requires the use of data sets made of \textit{ab initio} energies and forces. We choose a simple and general method to generate these data, making several small (important for modes with high frequencies) and large (important for modes with low frequencies) displacement of each atom from the equilibrium position in the three Cartesian directions.
With a large number of displacements, the PES is well described. However, the computing time increases with the number of displacements. It will therefore be necessary to find a good compromise between precision and number of geometries. So far, our results were obtained with $3N\times8$ displacements. To study the effect of the number of data points, we calculated the fundamental frequencies using a ANN potential, fitted on $3N\times6$ displacements ( $\pm0.01$, $\pm0.05$, $\pm0.2$ \AA), by removing the very large ones. 
First, for the neural network optimization, we obtained a RMSD about 9.2 meV/\AA\, for forces and 0.11 meV for energies, which is smaller than what was obtained with a larger data set (14.8 meV/\AA\, for forces and 0.21 meV for energies). This is simply because there are fewer data points to fit. Then, the RMSD between the full B2PLYP fundamental frequencies and the Hybrid ones is about 24.5 cm$^{-1}$ to be compared to 21.0 cm$^{-1}$ observed using the $3N\times8$ geometries. The result with the smaller data sets is slightly less accurate than that of the larger ones but the effect is not significantly important. By removing geometries with very large displacements ($0.3$\AA\,), the error for modes with lowest frequenies grows, especially for \ce{CH2NH} and \ce{C3H4O3} molecules, which explains the increase in the value of RMSD. To explain this overall influence of the dataset size, the correlation between the fitting error and the number of data points depends on the number of free parameters in the descriptor space. In the case of the implemented ANN, the number of free parameters is not much lower than the number of data points. As such, fitting with too few points leads to over fitting issues and we obtain better accuracy within the learning set but worse outside of it. In addition, going from $3N \times 8$ to $3N \times 6$ data points was achieved by removing the largest displacements which turns out to be crucial. Indeed, low-frequency modes require a large displacement and it has been mentioned in the literature that truncature to the fourth order of the polynomial representation of the PES is not sufficient for a correct representation of the  vibrational motion  of theses modes\cite{Inostroza2011}.  We note that by adding large displacements, the CPU computing time is increased by a factor of 8/6, but it does not necessarily increase the real computing time because the calculations of energies and forces are independent and can be done in parallel.

\section{Conclusion}

To summarize, anharmonic contributions to vibrational frequencies are usually either computed with empirical correction factors or with a full quantum mechanics calculations. We proposed an alternative approach which combines quantum chemistry calculations with machine-learning potential. In particular, B2PLYP/def2tzvpp is employed to compute equilibrium geometries and harmonic contributions while the anharmonicity is obtained through the GVPT2 framework where derivatives of the PES are computed with a neural network force-field potential. With this approach, we managed to reach a RMSD equal to 23\,cm$^{-1}$ when compared to experimental results and to 21\,cm$^{-1}$ when compared to full B2PLYP/def2tzvpp calculations. Yet, we note that in few cases the deviation can reach up to $100$\,cm$^{-1}$. Further refinements regarding the size of the database or the neural network methodology should help fixing those outlying results. In terms of computational timing, we gained a factor of almost ten with the latter method which allows us to study large molecules. Moreover, the use of a neural network force-field makes our approach transferable to molecules made of any types of atoms. Furthermore, we demonstrate that the PES obtained with machine-learning is accurate enough to compute anharmonic frequencies. We implemented the approach within the GVPT2 framework, but the obtained PES could also be employed with other vibrational frequency methods such as VSCF, cc-VSCF, VCI-P. Finally, beyond the calculation of vibrational frequencies, our work participates in the common effort to speed up ab initio calculations and is therefore an additional evidence that machine-learning technique is a very effective tool for material science. 


\section{Acknowledgement}
\label{Acknowledgments}

This work was only possible thanks to the generous grants of computer time by P2CHPD (Universit\'e Lyon 1) and the "Centre de calcul CC-IN2P3" at Villeurbanne, France. JL acknowledges financial support of the Fonds de la Recherche Scientifique - FNRS. 

\section{Supporting Information Available}

File available free of charge : 
\begin{itemize}
  \item SuppInfo.pdf: List of all calculated frequencies, Group symmetry functions and the statestical parameters of machine learning study.
\end{itemize}

%


%
\nocite{*}

\end{document}